\documentclass{aastex63}

\usepackage{lineno}
\usepackage[utf8]{inputenc}
\usepackage[titletoc, title]{appendix}
\usepackage{CJKutf8}
\usepackage{float}
\usepackage{blindtext}
\usepackage{gensymb}
\usepackage{amsmath}

\newcommand{\be}{\begin{equation}}
\newcommand{\ee}{\end{equation}}
\def\lta{\,\raise 0.3 ex\hbox{$ < $}\kern -0.75 em
 \lower 0.7 ex\hbox{$\sim$}\,}
\def\gta{\,\raise 0.3 ex\hbox{$ > $}\kern -0.75 em
 \lower 0.7 ex\hbox{$\sim$}\,}

\newcommand{\focus}{f_{\rm g}}

\def \Physics {Department of Physics, University of Michigan, Ann Arbor, MI 48109, USA}
\def \Astronomy {Department of Astronomy, University of Michigan, Ann Arbor, MI 48109, USA}

\graphicspath{{Figures/}}

\begin{document}
\shorttitle{Removal of Earth's Trojan Asteroids}
\shortauthors{Napier et al.}

\title{A Collision Mechanism for the Removal of Earth's Trojan Asteroids}

\correspondingauthor{Kevin J. Napier}
\email{kjnapier@umich.edu}

\author[0000-0003-4827-5049]{Kevin~J.~Napier}
\affiliation{\Physics}
\author[0000-0002-2486-1118]{Larissa~Markwardt}
\affiliation{\Astronomy}
\author[0000-0002-8167-1767]{Fred C.~Adams}
\affiliation{\Physics}
\affiliation{\Astronomy}
\author[0000-0001-6942-2736]{David~W.~Gerdes}
\affiliation{\Physics}
\affiliation{\Astronomy}
\author[0000-0001-7737-6784]{Hsing~Wen~Lin (\begin{CJK*}{UTF8}{gbsn}
林省文\end{CJK*})}
\affiliation{\Physics}

\begin{abstract}
    Due to their strong resonances with their host planet, Trojan asteroids can remain in stable orbits for billions of years. As a result, they are powerful probes for constraining the dynamical and chemical history of the solar system. Although we have detected thousands of Jupiter Trojans and dozens of Neptune Trojans, there are currently no known long-term stable Earth Trojans. Dynamical simulations show that the parameter space for stable Earth Trojans in substantial, so their apparent absence poses a mystery. This work uses a large ensemble of N-body simulations to explore how the Trojan population dynamically responds if Earth suffers large collisions, such as those thought to have occurred to form the Moon and/or to have given Earth its Late Veneer. We show that such collisions can be highly disruptive to the primordial Trojan population, and could have eliminated it altogether. More specifically, if Earth acquired the final 1\% of its mass through ${\cal O}(10)$ collisions, then only $\sim1\%$ of the previously bound Trojan population would remain.  
\end{abstract}


\submitjournal{The Planetary Science Journal}
\accepted{20 April, 2022}

\section{Introduction}

The restricted circular three-body problem predicts five points of equilibrium called the Lagrange points. Three of these points (L1, L2, and L3) are colinear with the primary and secondary masses, and the remaining two (L4 and L5) lead and trail the secondary mass by 60 degrees respectively. The three colinear points are all saddle points of the potential, and are therefore unstable. However, L4 and L5 are points of linearly stable equilibrium (as long as the secondary/primary mass ratio $\mu < 0.039$, see \citealt{MurrayDermott}), and as such can host long-term stable populations colloquially known as Trojan asteroids. 

Because Trojan asteroids are in the 1:1 mean-motion resonance with their host planet, they can remain stable for billions of years. As a result, they have the potential to be of primordial origin and can provide important probes of the solar system's dynamical and chemical history. The Jupiter Trojans are of considerable interest, as their quantity and chemical composition may encode sensitive information about the composition of the protoplanetary disk and the migration history of the giant planets. Understanding this population is of such high importance that NASA's Lucy spacecraft will fly by seven Jupiter Trojans over the next decade, four of which are in binary systems \citep{Levison2021}. Neptune Trojans encode similar information and are now being studied in detail, with characterization efforts including a Cycle 1 James Webb Space Telescope program to obtain near-infrared reflectance spectra \citep{LarissaJWST}.

Only two Earth Trojans (ETs) have been found to date, and both are transient; simulations show that they are stable over time scales much shorter than the age of the solar system \citep{Connors2011, Hui2021}. The prospect of discovering long-term stable ETs is enticing, as their proximity and low delta-$v$ with respect to Earth  would make a spacecraft mission imminently possible. 
Additionally, it would likely be possible to return samples of ETs. This feature would be extremely valuable for studying the composition of the protoplanetary disk if the objects are indeed of primordial origin. If such Trojans are not of primordial origin, the samples would still be useful for gleaning insight about the reservoirs that populate near-Earth orbits. In light of the potential scientific payoff of discovering ETs, several surveys have been dedicated to searching for such objects (\citealt{Whitely1998, Cambioni2018, Yoshikawa2018, Larissa2020, Lifset2021}). While these searches have failed to discover any long-term stable ETs, they have placed significant observational upper limits on the possible population. Specifically, the number of ETs with sizes larger than approximately 400 meters is limited to at most 100 objects \citep{Larissa2020, Lifset2021}.

These observations are seemingly in tension with numerical simulations showing that ETs can be stable over Gyr timescales (see, e.g., \citealt{Tabachnik2000, Cuk2012, Marzari2013, Zhou2019, Malhotra2019, Christou2021}). Furthermore, while it has been shown that the Yarkovsky effect \citep{Bottke2006, Marzari2013, Zhou2019} can remove the small ETs (smaller than tens of kilometers) from resonance, the effect is not strong enough to remove the larger bodies on time scales comparable to the age of the solar system. Current theories therefore suggest that long-term stable ETs can exist. To date, any such object has eluded discovery.

Although the existence of any large ETs has not been ruled out by observational surveys, the continued absence of even a single long-term stable ET merits the consideration of a gap in our understanding of their dynamics. In this work we consider how collisions between asteroids and the Earth, such as those hypothesized to have occurred during the Moon-forming impact (see, e.g., \citealt{Canup2012, CukMoon}) or while Earth was acquiring its Late Veneer (see, e.g., \citealt{Thorsten2011, Raymond2013, Brasser2016}), would affect an initially-stable ET population. In Section \ref{sec:theory} we derive the circumstances of the collisions between the Earth and the asteroids that may have been responsible for delivering its Late Veneer. In Section \ref{sec:simulations}, we describe our numerical experiments in which we simulated the bombardment of Earth by large asteroids. In Section \ref{sec:analysis} we present the results of our simulations and analyze their outcomes. In Section \ref{sec:conclusions}, we summarize our results and interpret them in the broad context of the possible Earth Trojan population. For completeness, Appendix \ref{sec:corrections} provides a quantitative assessment of some of our approximations and Appendix \ref{sec:etremove} presents an order of magnitude estimate for the orbital changes required to remove Trojans from bound states. Finally, we consider perturbations of Earth's orbit due to close encounters that do not result in collisions (Appendix \ref{sec:close}) and direct collisions between incoming asteroids and Trojans (Appendix \ref{sec:trojancollide}).


\section{Analytic Theory}
\label{sec:theory}

In this section we compute the distributions of relative velocity and orientation for potential Earth impactors (i.e., asteroids on Earth-crossing orbits). We use the resulting distributions to compute the change in Earth's momentum for our simulations of asteroid-Earth collisions in Section \ref{sec:simulations}.

\subsection{Scales and Ordering}

Consider an asteroid on an Earth-crossing orbit. Here the asteroid has mass $m$, the Earth has mass $M_{\Earth}$, and the Sun has mass $M_{\odot}$. The masses obey the ordering
\begin{equation}
    m \ll M_{\Earth} \ll M_{\odot}\,,
\end{equation}
so that the orbits can be considered in the hierarchical limit. Using the mass of Earth, the mass ratio $\mu = M_{\Earth}/M_{\odot} \sim 3 \times 10^{-6}$.

As a first approximation, we take Earth to have a circular orbit with semi-major axis $a_\earth = 1$ au. The potential impactor has initial orbital elements ($a$, $e$) and is considered to have the same orbital plane as Earth ($i = 0$). The Hill sphere of Earth is given by
\begin{equation}
    R_H = a_\earth \left( \frac{M_\earth}{3 M_\sun}\right)^{1/3} \approx 0.01 \text{ au} ,
    \label{eq:hill}
\end{equation}
whereas the Sphere of Influence has the form
\begin{equation}
    R_S = a_\earth \left( \frac{M_\earth}{M_\sun}\right)^{2/5} \approx 0.006 \text{ au}.
    \label{eq:soi}
\end{equation}
Since the orbital length scales obey the ordering 
$$ R_S \sim R_H \ll a_\earth < a$$ we can use intermediate asymptotics to describe the dynamics of the bodies that collide with Earth. In other words, the orbit of the asteroid has a Keplerian orbit around the Sun until it reaches the vicinity of Earth. Once the asteroid enters the Earth’s sphere of influence, its subsequent dynamics are governed by the gravitational potential of Earth. We can thus find the velocity vector of the asteroid at the point where it crosses the orbit of Earth. This velocity vector then provides the outer boundary condition for the inner problem where the asteroid collides or interacts with Earth.

\subsection{Relative Velocity for Earth Encounters}
We now calculate the speed of the asteroid with respect to Earth at the point of orbit crossing. The asteroid orbit has specific energy and angular momentum given by
\begin{equation}
    E = -\frac{GM_\sun}{2a} \qquad \text{and} \qquad J^2 = GM_\sun a (1 - e^2)
\end{equation}
The azimuthal velocity component of the asteroid orbit is determined by conservation of angular momentum so that 
\begin{equation}
    v_{\phi\sun}^2 = \frac{J^2}{r^2} = \frac{G M_\sun a}{r^2}(1 - e^2)
\end{equation}
Note that the azimuthal velocity is defined here in the solar reference frame, as indicated by the subscript. The radial velocity of the asteroid with respect to the Sun is then given by
\begin{equation}
    v_r^2 = 2E - \frac{J^2}{r^2} + \frac{2 G M_\sun}{r} = \frac{2 G M_\sun}{r} - \frac{G M_\sun a}{r^2}(1 - e^2) - \frac{GM_\sun}{a}.
\end{equation}
Because Earth is assumed to have a circular orbit, it has no radial velocity component, so that the radial velocity of the asteroid is the same in both reference frames. In this approximation, the speed of the asteroid with respect to Earth is thus given by
\begin{equation}
    v_\infty^2 = v_r^2 + (v_{\phi\sun} - v_\earth)^2 = v_r^2 + v_{\phi\sun}^2 + v_\earth^2 - 2 v_{\phi\sun} v_\earth
    \label{eq:v_inf}
\end{equation}
where $v_\earth = (G M_\sun/a_\earth)^{1/2} \approx 30$ km/s is the orbital speed of Earth, and where all quantities are evaluated at $r = a_\earth$.\footnote{We have assumed in this calculation that the impactors are on prograde, bound orbits. If the orbits are bound and retrograde, then the minus sign in Equation (\ref{eq:v_inf}) becomes a plus sign.} Using the results obtained above we find
\begin{equation}
    v_\infty^2 = \frac{G M_\sun}{a_\earth} \left[ 3 - \frac{a_\earth}{a} - 2\left(\frac{a}{a_\earth}\right)^{1/2} (1 - e^2)^{1/2}\right].
\end{equation}
We see that the orbital speed of Earth $v_\earth$ sets the velocity scale for this collision problem. Let us define the dimensionless variables 
\begin{equation}
    u_\infty \equiv \frac{v_\infty}{v_\earth} \qquad \text{and} \qquad \xi \equiv \frac{a}{a_\earth}
\end{equation}
so that the expression for the hyperbolic speed at infinity reduces to the form
\begin{equation}
    u_\infty^2 = 3 - \frac{1}{\xi} - 2\xi^{1/2}(1-e^2)^{1/2}
\end{equation}
The largest possible speed is thus $u_\infty = \sqrt{3}$ or $v_\infty = v_\earth \sqrt{3} \approx 52$ km/s.\footnote{Since we have assumed bound orbits, the incoming asteroids travel on elliptical orbits. In principle, incoming objects from outside the solar system could encounter Earth on hyperbolic orbits. This would allow for larger encounter speeds. Notice also that these types of interactions could either take place in the solar birth cluster or in the field. Such encounters are expected to be rare in the field, but could be more common in the birth cluster (see also \citealt{Napier2021}).} This case corresponds to a distant asteroid orbit (large $a$) with small perihelion (eccentricity $e$ close to unity) so that the asteroid velocity is nearly radial when it encounters Earth. For more typical cases, we expect perihelion $q = \xi (1 - e) \sim 1$ in dimensionless units, and $\xi \sim 2$, so that $u_\infty \sim 0.2$ and $v_\infty \sim 6$ km/s. Note that the expressions derived above also apply for orbits with $a < a_\earth$ (although such objects do not originate from the asteroid belt). 

\subsection{Distributions of Impact Variables}

Using the results derived above, we can find the distribution of relative speeds by assuming a simple model for the initial population of asteroid impactors. We can also determine the distribution of angles at which the asteroids intercept the orbit of Earth. A great deal of previous work has quantified the distributions of speeds (and other properties) for near Earth asteroids for purposes of planetary defense (e.g., see \citealt{Bottke2000,Jeffers2001,Steel1998}; and many others). These distributions are consistent with the model developed below (although the present-day and primordial distributions could be different).

\begin{figure}
    \centering
    \includegraphics[width=0.95\textwidth]{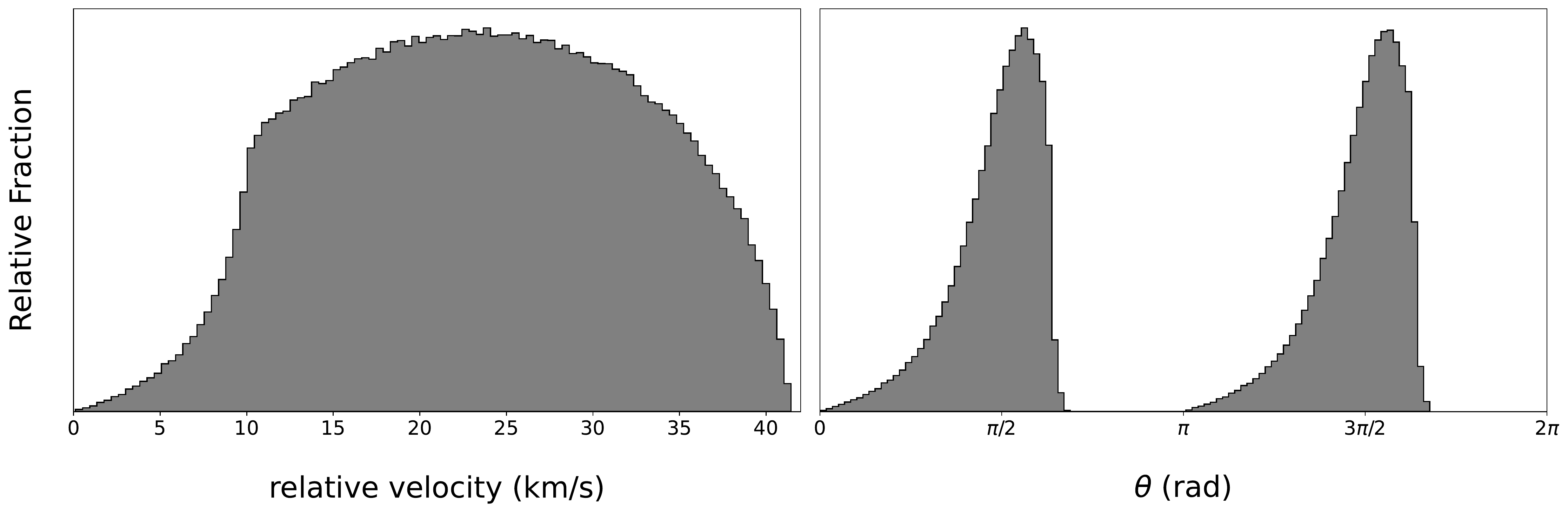}
    \caption{Distributions of incoming speeds (left) and incoming direction (right) for asteroids with orbits leading to close encounters with Earth. In both plots, the asteroid orbits have semi-major axes in the range $a =$ 1--5 au, and a uniform distribution of eccentricity. Only those asteroid orbits with sufficiently large values of eccentricity e cross the orbit of Earth and are included in the distribution. The two peaks in the distribution of incoming direction are a result of the impacting bodies being equally likely to hit Earth on the incoming or outgoing trajectories.}
    \label{fig:distributions}
\end{figure}

Let us consider the case where the asteroids have a surface density distribution of the form $\sigma \propto 1/a$, so that the distribution of semi-major axes of the asteroids is uniformly distributed within some range, i.e.,
\begin{equation}
    \frac{dN}{da} = constant \qquad \text{for} \qquad a_1 \leq a \leq a_2.
\end{equation}
For the sake of definiteness, here we assume $a_1 = 1$ au and $a_2 = 5$ au, so we are sampling the full annulus between the current orbits of Earth and Jupiter. For a given value of $a$, the eccentricity must be larger than a minimum value given by
\begin{equation}
    e_{min} = 1 - \frac{a_\earth}{a}.
\end{equation}
If this condition is not met, then the asteroid cannot cross the orbit of Earth. The eccentricity is unlikely to be much larger than that corresponding to an Earth crossing orbit---it takes interactions to pump up the eccentricity, and continued orbit crossings will eventually lead to large orbital perturbations. For the sake of definiteness, let us assume that the maximum eccentricity corresponds to perihelion of $a_\earth / 10$ so that
\begin{equation}
    e_{max} = 1 - \frac{a_\earth}{10 a}
\end{equation}
We then assume that the distribution of eccentricities is uniform over the interval $[e_{min}, e_{max}]$. With these choices for the distributions of the asteroid orbital elements ($a$, $e$), the resulting distribution of relative speeds $v_\infty$ is shown in Figure \ref{fig:distributions}.


We can also find the distribution of directions. Let $\theta$ be the angle of the incoming velocity (in the planetary reference frame) with respect to the velocity of Earth. In other words,
\begin{equation}
    \cos \theta = \frac{\mathbf{v} \cdot \mathbf{v_\earth}}{|\mathbf{v}| |\mathbf{v_\earth}|} = \frac{(v_{\phi\sun} - v_\earth)}{v_\infty}.
\end{equation}
The resulting distributions of both relative speed and angular direction are shown in Figure \ref{fig:distributions}. These distributions are used as the starting point for the numerical simulations discussed in the following section. Note that the distribution of angles has two peaks, corresponding to bodies impacting Earth on incoming and outgoing trajectories. Moreover, the two parts of distribution, centered on $\theta=\pi/2$ and $3\pi/2$, each include both positive and negative values of $\cos\theta$. In Appendix \ref{sec:corrections} we investigate the validity of the approximations made in this section.

\section{Numerical Simulations}
\label{sec:simulations}

The primary contribution of this work is an ensemble of numerical experiments where we tested the stability of a synthetic population of long-term stable ETs while Earth was stochastically bombarded by asteroids. For all of our numerical experiments, we used \texttt{Rebound's WHFAST} integrator \citep{reboundwhfast} and a model of the solar system that included the Sun, Venus, Earth, Mars, Jupiter, Saturn, Uranus, and Neptune.\footnote{We found that our results were not changed by resolving the Earth-Moon system. In the interest of saving computation time, we chose to use the Earth-Moon barycenter.} 

\begin{figure}[h]
    \centering
    \includegraphics[width=0.4\textwidth]{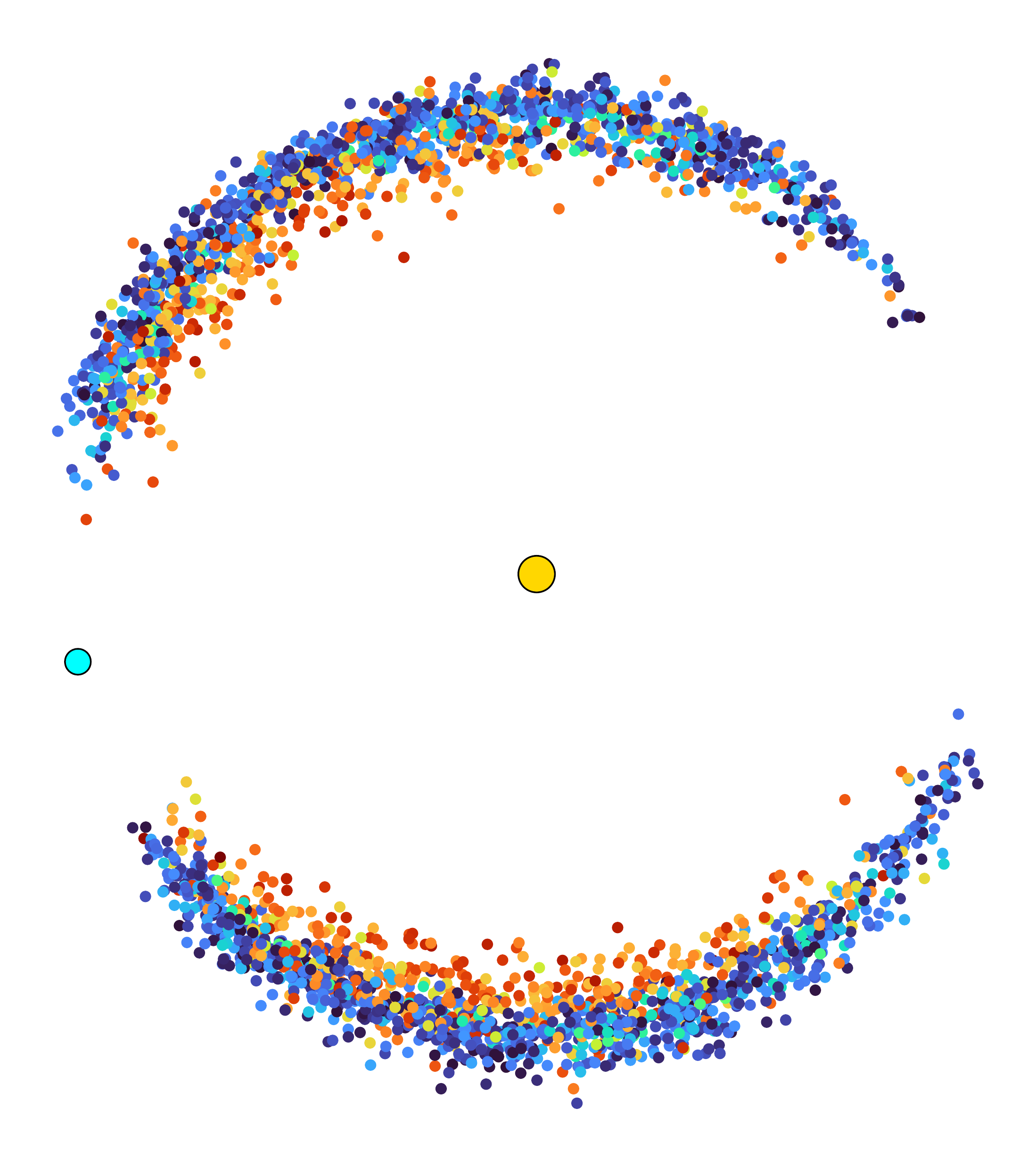}
    \includegraphics[width=0.45\textwidth]{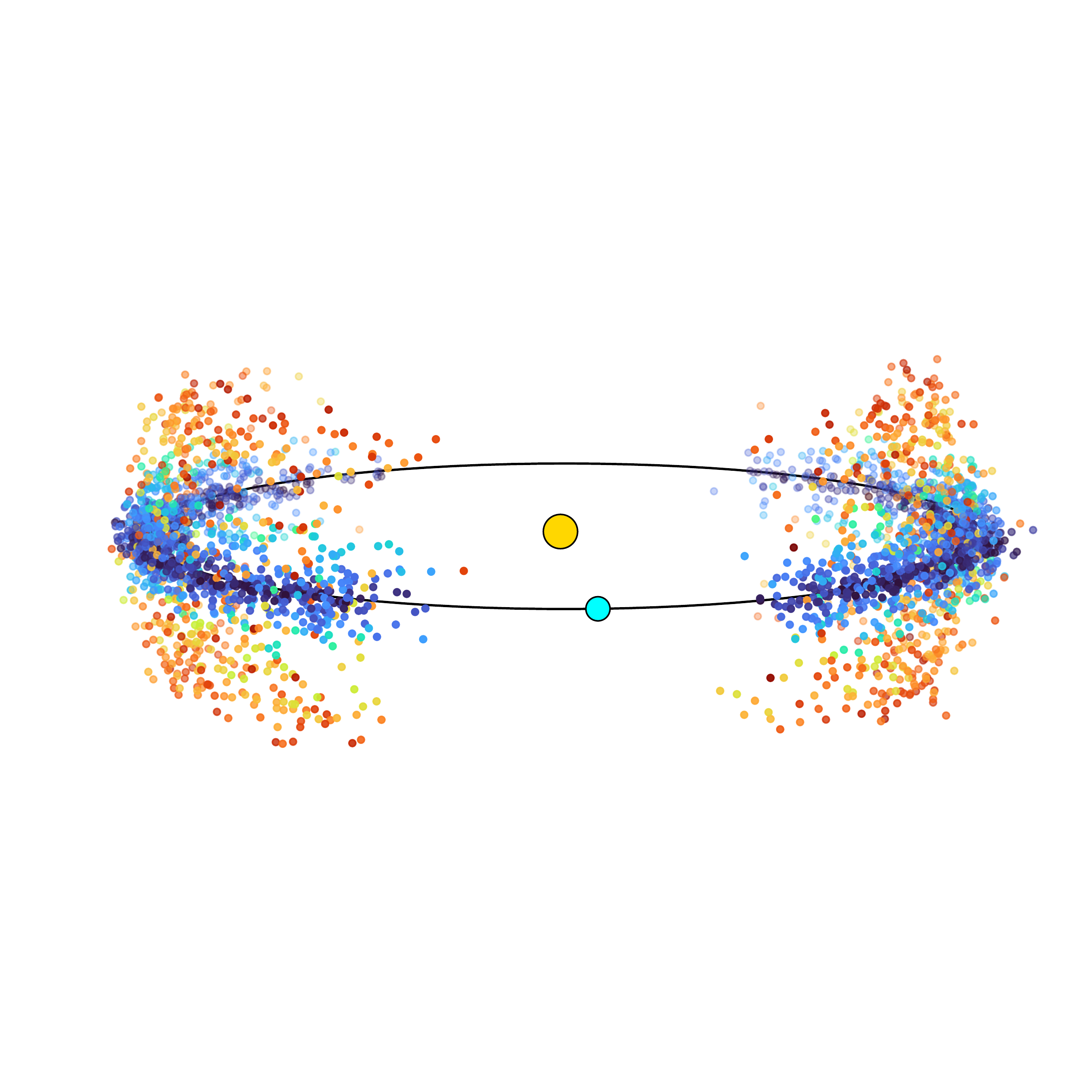}
    \caption{Our synthetic population projected into the ecliptic plane (left) and viewed from a higher inclination (right). The large cyan circle represents Earth, the large yellow circle represents the Sun, and the test particles are represented by smaller circles with colors corresponding to their inclinations (blue represents low inclination, and red represents high inclination, maxing out at approximately $40\degree$).}
    \label{fig:cloud}
\end{figure}

We began by generating a synthetic population of 3,021 ETs, each of which was dynamically stable for at least 10 Myr. This population represents the stable subset of an initial set of candidate ETs (which were integrated numerically to verify stability). The surviving objects were split roughly evenly between L4 and L5. We show the resulting population projected into the ecliptic plane in Figure \ref{fig:cloud}. The dynamical properties of our synthetic population match those reported by \citet{Zhou2019}. Note that due to the presence of other planets, these stable clouds are wider than one would expect by considering only the circular restricted three body problem (e.g., \citealt{MurrayDermott}).

Next we ran simulations to test the stability of our synthetic ETs as Earth underwent collisions with asteroids. We began each simulation by setting Earth's mass to 0.99 $M_\earth$, and integrating for a warm-up period of 100 years with no collisions. After the warm-up period, we perturbed Earth stochastically by modifying its velocity as it would change due to a collision with an asteroid.\footnote{For simplicity, we assumed a totally inelastic collision so that the entire mass of the impactor was added to Earth's mass. The velocity perturbations were calculated such that linear momentum was conserved, and were applied instantaneously.} We randomly drew the impactor velocity and impact angle from the distributions shown in Figure \ref{fig:distributions}. We chose the impactor mass by drawing a radius $R$ from a distribution of the form $N(R) \propto R^{-2}$ with values ranging from $500$ km to $2000$ km, and assuming a mass density $\rho = 3$ g cm$^{-3}$. While there are certainly many asteroids smaller than 500 km, they will deliver a negligible amount of mass to Earth compared to the mass delivered by a few large objects. Once we had perturbed Earth, we integrated for 10,000 more years (we found that this was enough time for the system to roughly settle back into equilibrium), tracking whether each test particle remained a Trojan, migrated onto a horseshoe orbit, or was ejected from a co-orbital state altogether. If an object was no longer a co-orbital with Earth, we removed it from the simulation. We continued this procedure until Earth's mass reached 1 $M_\earth$.\footnote{If a collision would have resulted in Earth's mass exceeding 1 $M_\earth$, we simply reduced the mass of the impactor such that Earth's mass became exactly 1 $M_\earth$.} We repeated this process a total of 5,000 times to build up our statistics. The numerical values used in these simulations were all chosen to be roughly consistent with the circumstances of Earth's Late Veneer impacts reported in the literature (see, e.g. \citealt{Raymond2013}, \citealt{Brasser2016}, and the references therein).

\section{Analysis}
\label{sec:analysis}

\begin{figure}[h]
    \centering
    \includegraphics[width=0.95\textwidth]{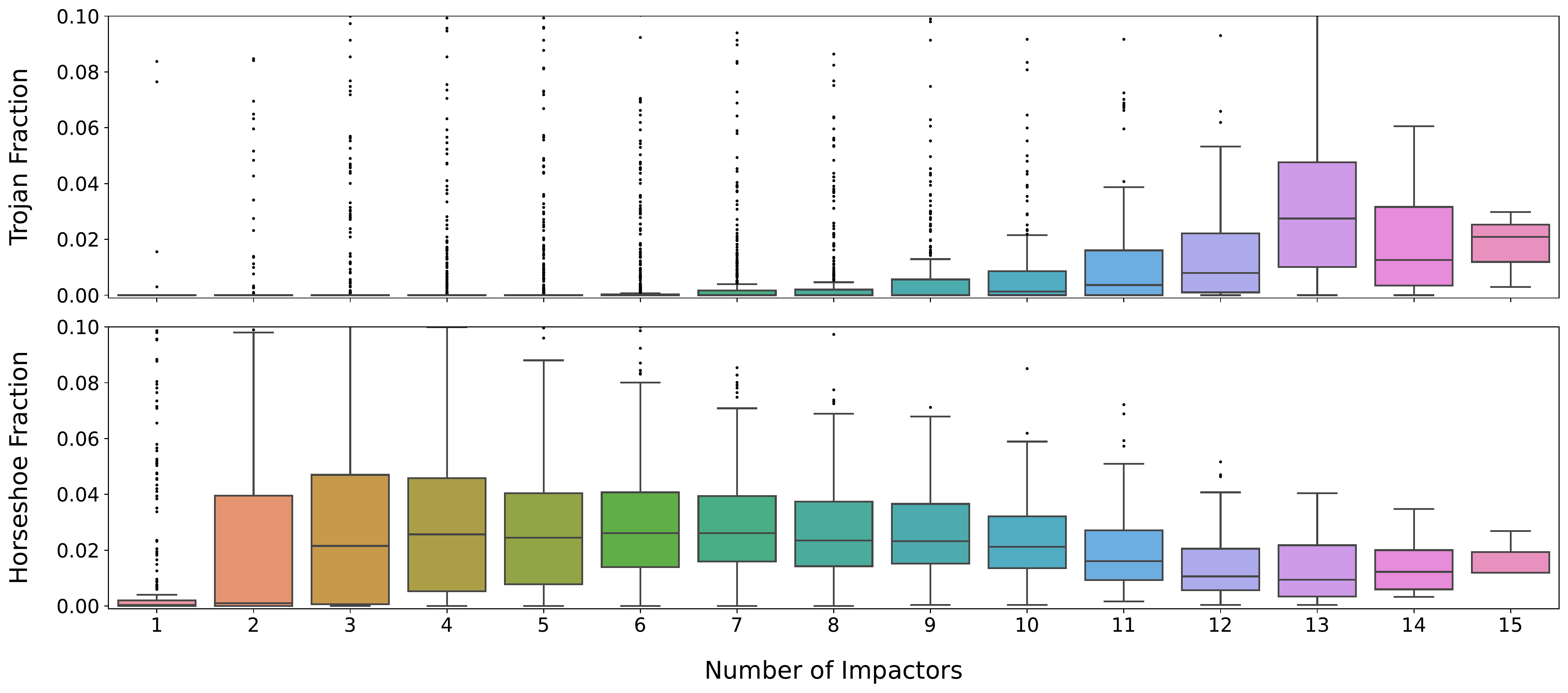}
    \caption{Box-and-whisker plots of the fraction of the initial population of 3,021 ETs that ended the simulation in a Trojan (top) or a horseshoe (bottom) orbit, plotted as a function of the number of impactors. There are data points that exceed the vertical range we have chosen for these plots, but such events are relatively rare.}
    \label{fig:survivors}
\end{figure}

In this section we analyze the data from our simulations. We begin by examining the probabilities with which initially-Trojan objects remained on true Trojan orbits, or were perturbed onto horseshoe orbits. In Figure \ref{fig:survivors}, we show box-and-whisker plots of the fraction of the initial population of 3,021 ETs that ended the simulation in a Trojan or a horseshoe orbit, as a function of the number of impactors. Note that there are data points that exceed the vertical range we have chosen for Figure \ref{fig:survivors}, but such events are relatively rare. After running many simulations an apparent trend emerges: for any number of impacts, the median surviving fraction of Trojans was less than a few percent. In addition, for a given number of impactors, the median number of objects surviving on horseshoe orbits was always on the order of 1-2 percent. These results indicate that large asteroid impacts with Earth are efficient at clearing the co-orbital population. 

\begin{figure}[h]
    \centering
    \includegraphics[width=0.95\textwidth]{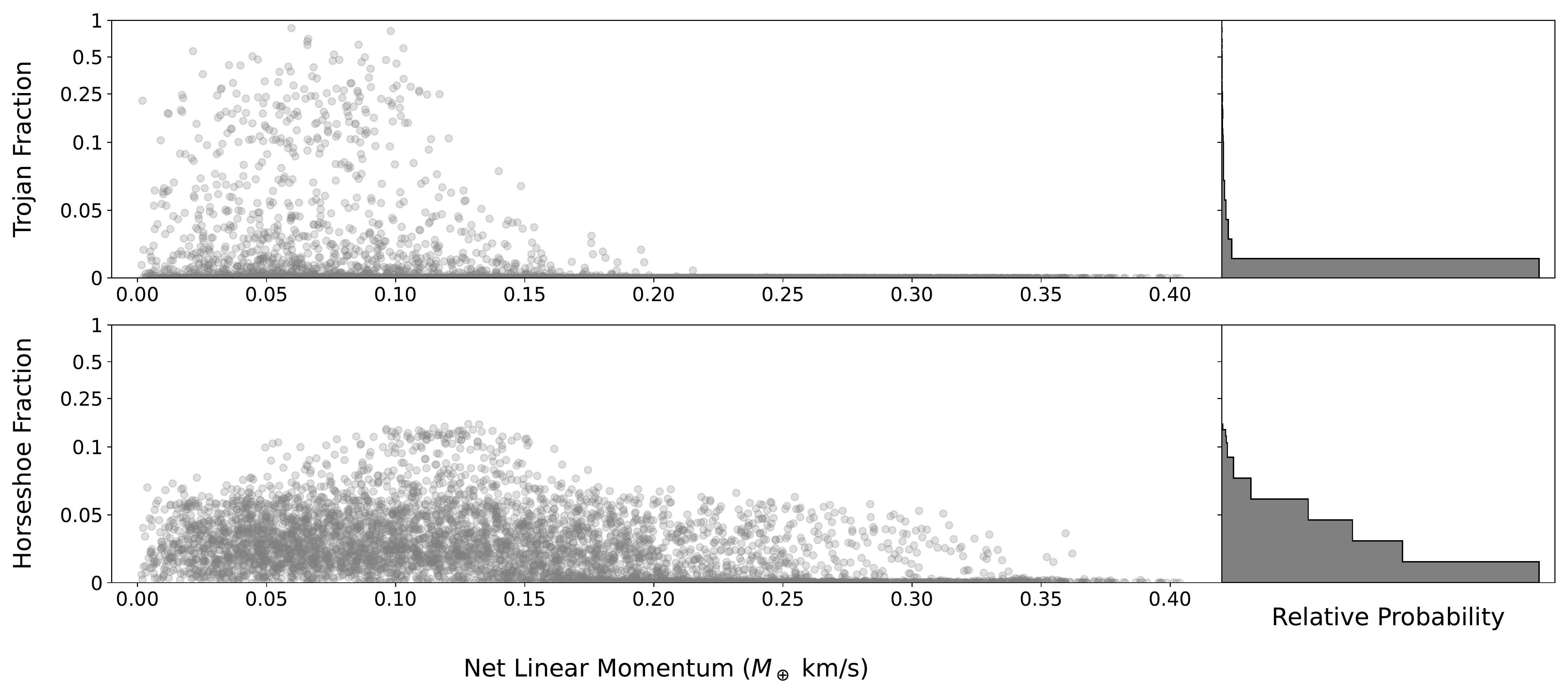}
    \caption{Scatter plot of the fraction of the initial population of 3,021 ETs that ended the simulation in a Trojan (top) or a horseshoe (bottom) orbit, plotted as a function of the net linear momentum of the impactors. The panels on the right-hand side are histograms representing the probability with which a given fraction of co-orbitals was retained. The fraction of objects remaining in co-orbitals decreases as the net linear momentum of the impactors increases. Note that the vertical axis is linear in the range $[0, 0.1]$ and logarithmic in the range $[0.1, 1]$.}
    \label{fig:survivors-vs-p}
\end{figure}

This series of collisions is a stochastic process in which different combinations of the number, velocities, and impact angles of the impactors can lead to vastly different outcomes. Indeed, there were some simulations in which an appreciable fraction of the objects remained on Trojan orbits, and as such require a more nuanced analysis. We find that all such simulations were instances where the number of impactors was small, and where all of them had low relative velocities. In these cases, the impactors simply did not impart enough momentum to Earth to remove the majority of the Trojan population. We demonstrate this trend in Figure \ref{fig:survivors-vs-p} by plotting the surviving fractions of Trojan and horseshoe orbits as a function of the net linear momentum of the impactors. As the net linear momentum imparted by impactors increases, the probability of many objects remaining on Trojan orbits decreases. As the number of impactors increases, the probability of drawing all impactors with low velocities decreases. This decreases the probability of a small net linear momentum, and thus decreases the frequency and severity of outliers. Of course, there is a limit to this logic. In the case that Earth gradually acquired its Late Veneer from millions of small rocks, the net linear momentum would converge to zero, and the Trojans would remain largely unaffected. In a similar vain, the most common outcome (by far) with 1--5 impactors was that no objects remained in Trojan orbits at the end of the simulation. This outcome occurs because with so few impactors, a large net linear momentum is far more probable than a small net linear momentum (see the distributions in Figure \ref{fig:distributions}).

\begin{figure}[h]
    \centering
    \includegraphics[width=0.95\textwidth]{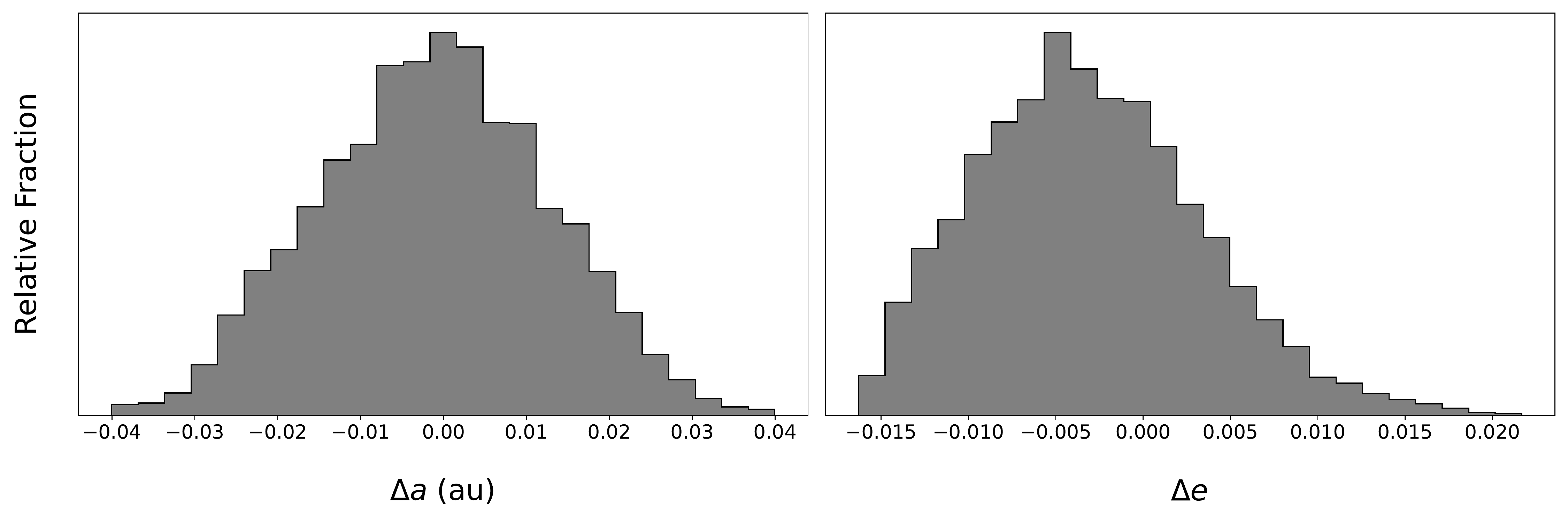}
    \caption{Histograms of changes in Earth's semi-major axis (left) and eccentricity (right) during each simulation. The initial values were $a = 1$ au and $e = 0.0167$.}
    \label{fig:earth-ae}
\end{figure}

Finally in Figure \ref{fig:earth-ae}, we show the distribution of changes in Earth's semi-major axis and eccentricity during our simulations. We find that Earth's semi-major axis never changes by more than a few percent, and its eccentricity remains reasonably small --- always within the range of predicted secular oscillations for the present-day solar system. The distribution of changes in semi-major axis is nearly symmetric about $\delta{a}=0$. In contrast, the distribution of changes in eccentricity is slightly skewed toward negative values (decreases in eccentricity, i.e. increases in angular momentum). Note that we began our simulations with Earth's $a = 1$ au and $e = 0.0167$. The exact starting values are not important for this study, as long as they are in the vicinity of today's observed values. If the starting eccentricity is close to zero, however, then the changes in eccentricity (while still small) must be skewed toward the positive. Finally, we note that our choice of orbital elements for the incoming rocks includes specific angular momenta that are both larger and smaller than that of Earth. Different choices of these distributions could lead to somewhat different changes in the orbit of Earth (see, e.g., \citealt{Fernandez1984}).

\section{Discussion and Conclusions}
\label{sec:conclusions}

In this paper we have examined the stability of the ETs in the face of collisions between Earth and massive asteroids. The main result of this work is the demonstration that large impacts to Earth can readily remove Trojans from resonance. Using an ensemble of 5,000 numerical simulations, we have demonstrated that if Earth acquired the final 1\% of its present-day mass from a handful of asteroidal impactors, the population of Earth Trojans would be reduced by more than $99\%$ (less than $1\%$ remaining; see Figure \ref{fig:survivors}). Due to the wide distribution of impactor parameters and the chaotic nature of the system, there are some instances in which a significant fraction of the original Trojans can remain on Trojan orbits. However, such instances are rather rare. In addition, we find that the population of objects on horseshoe orbits exceeds that of the Trojans, provided that the total number of collisions remains small. For the scenarios explored in this work, on average approximately 2\% of the primordial Trojan population ends up on horseshoe orbits.\footnote{As an important caveat, note that we did not integrate the objects for the age of the solar system, and therefore can not say whether they would survive until the present day.} As long as the number of impactors remains small, the number of Trojans expected to survive is at least an order of magnitude smaller than the horseshoe objects.\footnote{As the number of impactors increases, we expect to see more ETs and fewer horseshoe objects.} 

We also analyzed our simulation data for correlations between the impactor parameters and the surviving fraction of Trojans. First we find that as the number of impactors increases to more than of order $\sim10$, the surviving fraction of Trojans increases. This finding makes sense intuitively: If we consider the limiting case in which Earth acquired the final 1\% of its mass by having $10^{28}$ grains of sand slowly and isotropically scattered onto its surface, the resulting impulse would be a negligible and the Trojans would remain undisturbed. In a similar vein, we find a trend between the maximum retention fraction of the Trojan population and the net linear momentum imparted to Earth by the collisions (see Figure \ref{fig:survivors-vs-p}). As the net linear momentum increases, the upper bound on the surviving fraction of Trojans decreases. We found that no Trojans survived in any simulation with net linear momentum $\Delta {p}\gtrsim 0.25$ $M_\earth$ km/s (see also Appendix \ref{sec:etremove}). We also analyze the changes in Earth's semi-major axis and eccentricity at the end of the simulations to make sure that its orbit did not become unreasonably excited. The changes to the orbital elements are relatively small (see Figure \ref{fig:earth-ae}), with $\Delta{a}\sim0.01$ au and $\Delta{e}\sim0.005$.

This set of results has important implications for our understanding of the fate of primordial Earth Trojans. While numerical simulations show that stable Earth Trojans can exist, dedicated searches still have not discovered any such objects. It has been noted that the Yarkovsky effect  can remove small Trojans from resonance on the timescale of the age of the solar system, but it can not explain the apparent absence of large objects. In contrast, the collision mechanism explored in this paper is independent of the size of the Trojans. Moreover, this collision mechanism is of particular interest because previous work suggests that Earth has indeed experienced large impact events after its formation: one example being the impact that formed the Moon (e.g., \citealt{Canup2012}), and another a series of impacts in which Earth acquired its Late Veneer (e.g., \citealt{Raymond2013}). The Moon is thought to have formed as the result of a Mars-sized object colliding with the primordial Earth. Our simulations show that such a collision likely to have cleared out the primordial ETs (though perhaps not if the collision was relatively slow, thus imparting a rather small change in net linear momentum). However, a collision this large would probably have re-injected some material into Trojan resonances (note that similar ideas have been put forth in the context of Martian Trojans, e.g., \citealt{Polishook2017}). On the other hand, the collisions that gave Earth its Late Veneer are hypothesized to have occurred well after the Moon-forming collision. If the Late Veneer was comprised of a small number of large impactors, as  favored by recent models \citep{Raymond2013, Genda2017}, any re-injected material is likely to have been cleared. Furthermore, the impacts comprising the Late Veneer were probably not large enough to re-populate L4 and L5 with ejecta (especially bodies large enough to survive the Yarkovsky effect), but this complication should be explored in future work. Additionally, because the Late Veneer is thought to have occurred late in the formation of the inner solar system, there was probably not a large population of ambient bodies to repopulate the Trojan resonances \citep{Brasser2016}. Despite the large parameter space in which dynamically stable ETs can exist, it is possible that all or most such objects have been removed due to asteroids impacting Earth.

We note that this work has limitations and can be improved in the future. Given a considerable amount of additional computing, future studies should by more thoroughly explore the distributions of impactor parameters. In particular, changing the size, mass, and eccentricity distributions of the impactors could influence the results. For example, a shallower size distribution would mean that fewer impacts are necessary to deliver the required mass to Earth, while a steeper size distribution would mean that more impacts are necessary. In this vein, the total mass of the impactors is also important, and should be considered in future studies. We also made the approximation that the collisions were co-planar with Earth's orbit, while in reality they would have some distribution of inclinations. One should also address the possibility of impactors coming from outside of the asteroid belt. In this work we considered only impactors on orbits with $a \in [1, 5]$ au. Orbits with larger semi-major axes corresponding to the giant planet region or the Kuiper Belt would have marginally larger relative velocities when entering Earth's Hill sphere. Such collisions would deliver larger impulses to Earth than their low-$a$ counterparts, and would thus be more detrimental to the survival of the Trojan populations. However, this correction is rather small --- a body with $a \rightarrow \infty$ can achieve a maximum velocity of only $\sim 52$ km/s. We also did not model the effect of close encounters between Earth and potential impactors, or impacts directly onto the Trojan asteroids. We make analytic estimates of the relative sizes of these effects in Appendices \ref{sec:close} and \ref{sec:trojancollide} respectively. We find that the expected net velocity perturbation to Earth from close encounters is roughly equivalent to that expected from direct collisions. The perturbations from close encounters will be orthogonal to those from direct collisions, so including close encounters in the model should enhance the net perturbation to Earth by a factor of approximately $\sqrt{2}$. Direct collisions with Trojans, on the other hand, are a less likely source of orbital disruption than perturbations to Earth’s orbit by an order of magnitude. These considerations imply that the present work likely represents a lower limit on the disturbance to the Trojan population. Of course, the most important additional work is observational: Future observational surveys will either detect a (presumably small) population of Earth Trojans or place increasingly tight limits on their existence. With sufficiently tight constraints on the population of primordial ETs, future simulations can in turn constrain the properties of the primordial asteroid belt.

\appendix
\section{Velocity Corrections}
\label{sec:corrections}

In this section, we assess the severity of the approximations used in Section \ref{sec:theory}. In particular, we have assumed that the orbit of the asteroid only depends on the solar gravitational potential, and we have assumed that the orbit of Earth is circular.

In this approximation, we assume that the asteroid follows an elliptical orbit around the Sun, and then encounters Earth at the point where the orbits intersect. In this phase, the distance $d_\sun$ between the asteroid and Earth in the frame of the solar system thus vanishes ($d_\sun \rightarrow 0$) at the intersection. The approximation then switches over to a reference frame where the planet (Earth) is at rest, and the asteroid is assumed to have speed $v_\infty$ and distance $d_\sun \rightarrow \infty$. We are thus using the method of matched intermediate asymptotics, where the inner boundary condition for the outer problem ($d_\sun \rightarrow 0$) represents the outer boundary condition for the inner problem ($d_\sun \rightarrow \infty$). In order to test the consistency of this approach, we need to determine how much the gravitational potential of Earth affects the incoming speed $v_\infty$ calculated previously.

The boundary between the inner problem and the outer problem is delineated by the Hill sphere, as given by Equation (\ref{eq:hill}). At this Hill sphere boundary, the additional velocity that the asteroid would gain from the gravitational potential well of Earth is given approximately by
\begin{equation}
    v_x^2 = \frac{2 G M_\earth}{R_H} = \frac{2 G M_\earth}{a_\earth}\left( \frac{3M_\sun}{M_\earth} \right)^{1/3} = 3^{1/3} \frac{2 G M_\sun}{a_\earth} \left( \frac{M_\earth}{M_\sun}\right)^{2/3}
\end{equation}
As shown previously, the incoming speed $v_\infty$ is comparable to the orbital speed of Earth. If we let $v_T$ denote the true speed (corrected for Earth’s gravity), then
\begin{equation}
    v_T^2 = v_\infty^2 + v_x^2,
\end{equation}
which in turn implies
\begin{equation}
    \frac{v_T}{v_\infty} = \left[ 1 + \frac{v_x^2}{v_\infty^2}\right]^{1/2} \approx 1 + \frac{v_x^2}{2v_\infty^2}.
\end{equation}
The size of the correction is thus given by
\begin{equation}
    \frac{\Delta v}{v} = \frac{v_T - v_\infty}{v_\infty} \approx 3^{1/3}\left(\frac{M_\earth}{M_\sun}\right)^{2/3} \sim 0.0003.
\end{equation}
If we use the sphere of influence from Equation (\ref{eq:soi}), instead of the Hill radius, the above equation has the form
\begin{equation}
    \frac{\Delta v}{v} = \frac{v_T - v_\infty}{v_\infty} \approx \left(\frac{M_\earth}{M_\sun}\right)^{3/5} \sim 0.0005.
\end{equation}
The correction due to the gravity of Earth is thus relatively small.

Another correction arises from our assumption that Earth has a circular orbit. In this case, the planet has a constant speed $v_\earth^2 = GM_\sun/a_\earth$. For small eccentricity, the variation in speed over the course of the orbit is given approximately by the expression
\begin{equation}
    \left.\frac{\Delta v}{v}\right|_{max} = \left[ \frac{1+e}{1-e} \right]^{1/2} - 1 \approx e \approx 0.0167,
\end{equation}
where we have used the observed eccentricity of Earth. In summary, both corrections are small, less than about 1 percent. The correction for the eccentricity of Earth's orbit is larger than the correction due to the Earth’s gravity.

Next we consider the crossing time: The relative asteroid speeds are of order $v_\infty = 10$ km/s and the size of the Hill sphere is of order $R_H = 10^{11}$ cm. The crossing time of the asteroid is thus of order
\begin{equation}
    t_{cross} \approx \frac{R_H}{v_\infty} \sim 10^5 \text{ sec} \sim 1 \text{ day}.
\end{equation}
Over the crossing time ($\sim$ 1 year), the planet rotates through about $1/300$ of its orbit, so the reference frame of the planet does not rotate appreciably over the course of the encounter. We thus expect the effects of the rotating reference frame to be relatively small, but they can be quantified, as shown below.
The encounter takes place in a rotating frame of reference, with rotation rate $\Omega$ given by the mean motion of Earth. In a rotating reference frame, the acceleration is given by
\begin{equation}
    \mathbf{a}_R = \mathbf{a}_g - 2 \Omega \hat{z} \times \dot{\mathbf{r}} - \Omega^2\hat{z} \times (\hat{z} \times \mathbf{r}),
\end{equation}
where we have taken the angular velocity vector of the rotating frame to point in the $\hat{z}$ direction. On the right hand side of the equation, the first term is the acceleration due to the gravity of Earth, the second term is the Coriolis acceleration, and the third term is the centrifugal acceleration. In order of magnitude, we can find the relative size of the Coriolis term and the centrifugal term,
\begin{equation}
    \frac{a_{cor}}{a_g} = 2 \left( \frac{M_\sun r^3}{M_\earth a_\earth^3} \right)^{1/2} = \frac{2}{\sqrt{3}} \left(\frac{r}{R_H} \right)^{3/2} \qquad \text{and} \qquad \frac{a_{cen}}{a_g} = \left( \frac{M_\sun r^3}{M_\earth a_\earth^3} \right) = \frac{1}{3} \left( \frac{r}{R_H} \right)^3.
\end{equation}
At the Hill radius, both ratios are thus of order unity, but they both decrease rapidly as the asteroid penetrates the sphere and orbits closer to the planet. The centrifugal term is always smaller than the Coriolis term. We can thus estimate the size of this correction by using only the Coriolis term, and by noting that the change in velocity produced by this additional acceleration is given approximately by
\begin{equation}
    \Delta v \sim \int a_{cor} dt \sim \int 2\Omega v dt \sim 2\Omega dr < 2\Omega R_H = 2\left( \frac{GM_\sun}{a_\earth}\right)^{1/2}\left(\frac{M_\earth}{3M_\sun} \right)^{1/3}
\end{equation}
We thus obtain the approximate bound 
\begin{equation}
    \frac{\Delta v}{v} \lesssim 2\left(\frac{M_\earth}{3 M_\sun}\right)^{1/3} \approx 0.03
\end{equation}
If we use the sphere of influence instead of the Hill radius, the above expression becomes
\begin{equation}
    \frac{\Delta v}{v} \lesssim 2\left(\frac{M_\earth}{M_\sun}\right)^{2/5} \approx 0.01
\end{equation}
The correction for the rotating frame of reference is thus small, but it is somewhat larger than the corrections outlined earlier due to the influence of Earth in the solar reference frame and the non-circular nature of its orbit.

\section{Trojan Removal}
\label{sec:etremove} 

Here we make some rough estimates of the energetic requirements for the removal of Earth Trojans from bound orbits. In this context, a bound orbit is one such that the body is bound to either the L4 or L5 point, and hence executes a Trojan orbit. Unbound orbits are no longer Trojans, but still remain bound to the Sun.

We work in the approximation of the circular restricted three body problem, following the notation of \citet{MurrayDermott} and work in their dimensionless units where the mean motion of Earth is unity and the separation of the two majors bodies is also unity. In these units, the Jacobi constant has the form
\begin{equation}
    C_J = (x^2 + y^2) + 2 \left(\frac{1 - \mu}{r_1} - \frac{\mu}{r_2} \right) - v^2.
\end{equation}
The values of the Jacobi constants for the five Lagrangian equilibrium points are as follows:
\begin{align}
    C_1 &= 3 + 3^{4/3}\mu^{2/3} - 10\mu/3 \\
    C_2 &= 3 + 3^{4/3}\mu^{2/3} - 14\mu/3 \\
    C_3 &= 3 + \mu \\
    C_4 &= C_5 = 3 - \mu
\end{align}
We can thus define the difference in Jacobi constants between Lagrangian points:
\begin{equation}
    \Delta_{3-45} = C_3 - C_4 = 2\mu
\end{equation}
and
\begin{equation}
    \Delta_{12-45} = \frac{1}{2}(C_1 + C_2) - C_4 = (9\mu)^{2/3} - 3\mu
\end{equation}
As a rough approximation, the velocity increment required to move an object from a Trojan orbit to a horseshoe orbit is then given by
\begin{equation}
    \Delta v \sim \left( \Delta_{3-45}\right)^{1/2} v_\earth = (2\mu)^{1/2} v_\earth \approx 0.0024 v_\earth \qquad \implies \qquad \Delta v \sim 0.074 \text{ km/s}
\end{equation}
Similarly, the velocity increment required to make a Trojan become unbound from any libration is given approximately by
\begin{equation}
    \Delta v \sim \left( \Delta_{12-45}\right)^{1/2} v_\earth = (9\mu)^{1/3} v_\earth \approx 0.03 v_\earth \qquad \implies \qquad \Delta v \sim 0.9 \text{ km/s}
\end{equation}
The above values represent upper bounds on the velocity increments required to dislodge the Trojans. Note that when an impact takes place on Earth, the immediate collision changes the velocity of Earth, so that the velocity of the Trojan in the Earth-centric frame changes accordingly. However, there is another effect: the collision also induces an eccentricity in Earth's orbit, the magnitude of which will depend on the direction from which Earth is struck. As a result of the nonzero eccentricity, the speed of Earth relative to the Trojan continues to change over the course of an orbit, causing additional instability.

\section{Close Encounters} 
\label{sec:close}

The calculation of the main text shows that direct collisions between incoming rocky bodies and Earth can lead to the disruption of Trojan orbits. In addition to direct impacts, close passages of rocky bodies will also lead to orbital perturbations of Earth.  Since these close passages do not add mass to Earth, these encounters will enhance the  disruptive effects already considered (i.e., the disruption found in the main text should be considered as a lower limit). 
 
We can estimate the magnitude of this effect. Consider a field of rocky bodies that are streaming toward Earth. As a first approximation, we consider the rocks to have a given mass $m$, speed $v$ with respect to the Earth, and assume that all of them are coming from the same direction. One could generalize to distributions of $(m,v)$, but we start by comparing the velocity perturbations produced by direct collisions with those induced by close encounters.

The rate at which rocky bodies collide with Earth is given by 
\be
\Gamma = n v \pi \focus R_\oplus^2 \,,
\ee
where $n$ is the number density of rocks and where $\focus>1$ is gravitational focusing factor. It will be useful to work in terms of the number of impacts $N$, so we write 
\be
(\Delta t) = {N \over nv\pi \focus R_\oplus^2} \,. 
\label{deltat} 
\ee
If all of the collisions have the same sign, the velocity perturbation (to Earth) produced over time $(\Delta t)$ has the form 
\be
(\Delta v)_{\rm dir} \approx n v^2 \left({m\over M_\oplus}\right) 
\pi \focus R_\oplus^2 (\Delta t) = Nv \left({m\over M_\oplus}\right) \,. 
\ee
However, are interested in the square of the velocity perturbation. If the incoming rocks produce both positive and negative angular momentum perturbations, as expected, then the velocity perturbation grows as a random walk \citep{binneytremaine} so that
\be
(\Delta v)_{\rm dir}^2 \approx N v^2
\left({m\over M_\oplus}\right)^2  \,.
\ee
For completeness, we note that if the rocks all arrived from a single direction, then the expression would pick up an additional factor of $N$. 

Now consider the case of near-misses or close encounters. First we note that while the direct collisions produce velocity perturbations in the direction of motion of the rocky impactor, close encounters produce velocity perturbations in the perpendicular direction. Working in the impulse approximation, the velocity perturbation produced by a fly-by encounter \citep{binneytremaine} has the form
\be
\delta v = {2Gm \over bv} \,,
\ee
where $b$ is the impact parameter of the interaction (we suppress subscripts, but keep in mind that the direction is perpendicular). For close encounters, the interactions are equally likely to occur on either side of the planet, which means that the velocity perturbations have both positive and negative signs.  Moreover, the size of the perturbation depends on the impact parameter. The rate of encounters with impact parameter between $b$ and $b+db$ is given by
\be
\Gamma_b = n v 2\pi b db \,.
\ee
As a result, the square of the velocity perturbation accumulates according to the relation 
\be
d \left[{d \over dt} (\Delta v)_{\rm cl}^2 \right] 
= \left({2Gm\over b v}\right)^2
n v 2\pi b db \,.
\ee
After integrating over all impact parameters, and over time, we obtain  
\be
(\Delta v)_{\rm cl}^2 = 
{8\pi G^2 m^2 n (\Delta t) \over v} \ln\Lambda\,,
\ee
where 
\be
\Lambda \equiv {b_{\rm max}\over b_{\rm min}} \,. 
\ee 
The minimum impact parameter must be larger than Earth so that $b_{\rm min}\ge R_\oplus$. In general, we expect $b_{\rm min}$ $\approx \sqrt{\focus} R_\oplus$.  As a working approximation, we can take the maximum impact parameter to be the Hill radius, i.e.,  
\be
b_{\rm max} \approx R_H = 
a_\oplus \left({M_\oplus\over 3M_\odot}\right)^{1/3} \approx
0.01 \,{\rm AU}\,.
\ee
As result, the so-called Coulomb logarithm $\ln\Lambda\approx5$.
Removing the factor of $(\Delta t)$ in favor of $N$ we find 
\be
(\Delta v)_{\rm cl}^2 = 
{8 G^2 m^2 N \over v^2 R_\oplus^2 \focus} \ln\Lambda\,. 
\ee
Even though the perturbations are in the parallel and perpendicular directions, for direct and close encounter interactions respectively, we are nonetheless interested in their relative magnitude. Since the direct collisions are also subject to a random walk, the ratio of the two contributions is given by 
\be
{\cal R} = {(\Delta v)_{\rm cl}^2 \over (\Delta v)_{\rm dir}^2} = 
{8 \ln\Lambda \over \focus} \left({GM_\oplus\over R_\oplus v^2}\right)^2 
\approx {40 \over \focus} \left({v_{\rm esc} \over v}\right)^4 \,. 
\ee
The distribution of incoming speeds (Figure 1) shows a broad peak near $v\sim25$ km/s, whereas the escape speed of Earth $v_{\rm esc}\sim11$ km/s. The gravitational focusing factor $\focus=1 + (v_{\rm esc}/v)^2$, so that the ratio ${\cal R}\approx1$. In other words, in the absence of asymmetries in the collision stream, the velocity perturbations due to close encounters are roughly comparable to those induced by direct collisions. The effects considered in this paper thus represent a lower limit to the full degree of Trojan disruption. Moreover, since  the close encounter perturbations are perpendicular to the perturbations produced by direct collisions, the contributions add in quadrature, so we expect the enhancement factor to be $\sim\sqrt{2}$. 

\section{Collisions with Trojans} 
\label{sec:trojancollide} 

This Appendix considers another channel for the disruption of Trojan
orbits. Given the reservoir of rocky bodies on Earth-crossing orbits
that interact with Earth and provide the Late Vaneer, direct
collisions with the Trojans can occur. For a given Trojan, the
expected number of collisions is given by 
\be
N_T = n_T \sigma_T v (\Delta t) \,,
\ee
where $\sigma_T = \pi R_T^2$ is the cross section for the Trojan and
$n_T$ is the number density of rocky bodies large enough to disrupt 
the orbit. Eliminating $(\Delta t)$ using equation (\ref{deltat}), 
the above expression can be written in the form 
\be
N_T = N_\oplus {n_T \sigma_T \over n_\oplus \sigma_\oplus} \,,
\ee
where $n_\oplus$ is the number density of object considered in this
paper to provide perturbations of Earth's orbit. Note that as the size
$R_T$ of the Trojan decreases, the cross section $\sigma_T\sim R_T^2$
decreases, but the number density of objects $n_T$ large enough to
disrupt the body increases. Although the size distribution of 
Earth-crossing bodies in not well-known, the two effects roughly 
compensate. We can thus obtain an approximate result using
$n_T = n_\oplus$ and taking the cross section $\sigma_T$ to be 
that of an $R=1000$ km body. Using these values along with the 
size of Earth and the gravitational focusing factor, we find 
\be
N_T \approx {N_\oplus \over 50} \,. 
\ee
Here, $N_\oplus$ is the number of (large) impactors from the
simulations of the paper, where we find $N_\oplus=1-15$, with typical
values $\sim5-10$. This estimate suggests that direct collisions with
Trojans are a less likely source of orbital disruption than
perturbations to Earth's orbit, by a factor of 5 to 10. 

\acknowledgements

We thank Renu Malhotra for useful discussions, and two anonymous referees for useful comments. This material is based upon work supported by the National Aeronautics and Space Administration under grant No. NNX17AF21G issued through the SSO Planetary Astronomy Program and by the National Science Foundation under grant No. AST-2009096.

\bibliographystyle{aasjournal}
\bibliography{references}

\end{document}